\documentclass[aps,twocolumn,groupedaddress]{revtex4}
\usepackage{graphicx}
\usepackage{latexsym}
\usepackage{amssymb}
\usepackage{amsfonts}
\begin{document}
\title{Magnetic-domain-controlled vortex pinning in a
superconductor/ferromagnet bilayer}
\author{M. Lange}
\author{M. J. Van Bael}
\author{V. V. Moshchalkov}
\author{Y. Bruynseraede}

\affiliation{Laboratorium voor Vaste-Stoffysica en Magnetisme, K.
U. Leuven, Celestijnenlaan 200D, B-3001 Leuven, Belgium}

\date{\today}
\begin{abstract}
Vortex pinning in a type-II superconducting Pb film covering a
Co/Pt multilayer with perpendicular magnetic anisotropy is
investigated. Different stable magnetic domain patterns like band
and bubble domains can be created in the Co/Pt multilayer, clearly
influencing the vortex pinning in the superconducting Pb layer.
Most effective pinning is observed for the bubble domain state. We
demonstrate that the pinning properties of the
superconductor/ferromagnet bilayer can be controlled by tuning the
size, density and magnetization direction of the bubbles.
\end{abstract}

\maketitle
\begin{figure}
\includegraphics{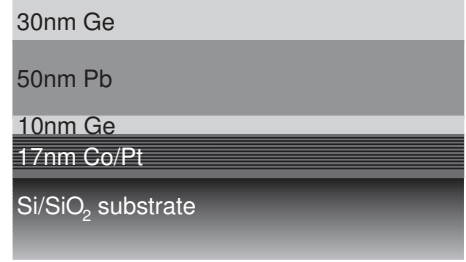}
\caption{Schematic presentation of the investigated
superconductor/ferromagnet layered structure.}
 \label{schema}
\end{figure}

\begin{figure}
\includegraphics{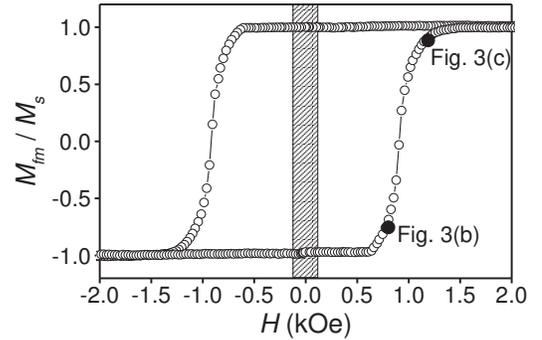}
\caption{Hysteresis loop of the Co/Pt multilayer measured by MOKE
at room temperature. Magnetic field $H$ is applied perpendicular
to the sample surface. The dashed area shows the field range where
magnetization measurements of the superconductor were carried out
(Fig.~\ref{mag}).}
 \label{moke}
\end{figure}

\begin{figure*}
\includegraphics{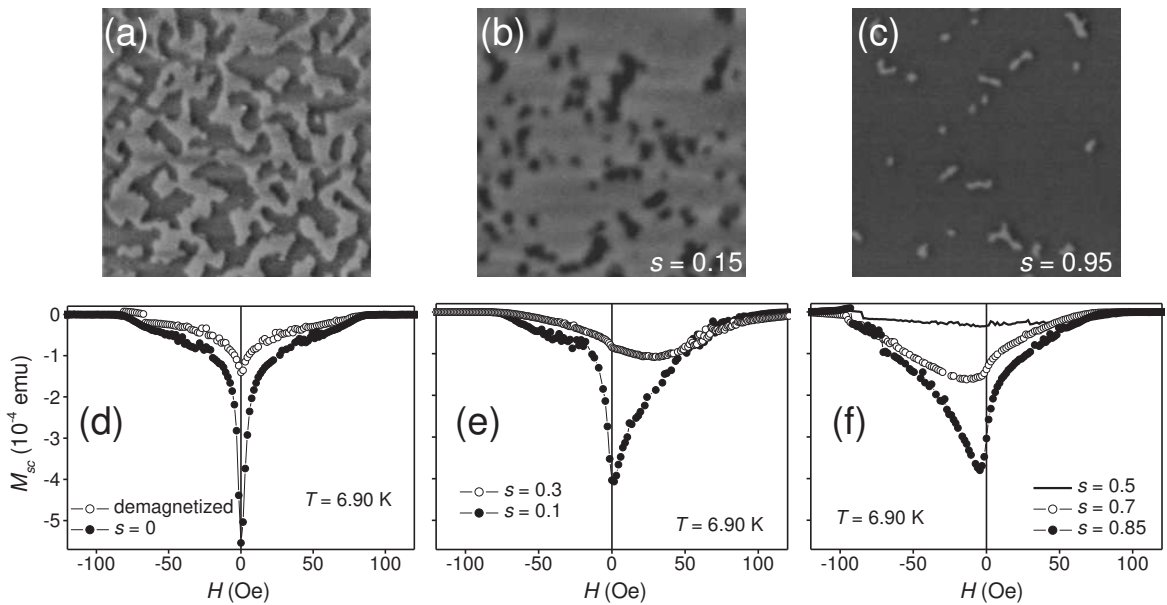}
\caption{(a)-(c)~MFM images (5~$\times$~5~$\mu$m$^{2}$) at room
temperature and $H=0$ of the Co/Pt multilayer in different
magnetic states: (a)~after demagnetization, (b)~after magnetizing
the sample in a perpendicularly applied field
$H=-10$~kOe~$\rightarrow +0.83$~kOe~$\rightarrow 0$ ($s=0.15$),
(c)~$H=-10$~kOe~$\rightarrow +1.10$~kOe~$\rightarrow 0$
($s=0.95$). (d)-(f)~Magnetization measurements of a type-II
superconducting Pb film covering a Co/Pt multilayer:
(d)~($\circ$)~after demagnetization, ($\bullet$)~in the $s=0$
state, (e)~($\circ$)~$s=0.3$, ($\bullet$)~$s=0.1$, and
(f)~($-$)~$s=0.5$, ($\circ$)~$s=0.7$, ($\bullet$)~$s=0.85$.}
 \label{mag}
\end{figure*}

 Magnetic flux penetrates type-II superconductors in the
form of quantized vortices, which have the tendency to form a
periodic lattice. These vortices move when a current is sent
through the superconductor, thus causing dissipation and limiting
the critical current density $j_{c}$ of the superconductor. To
enhance $j_{c}$, vortex motion must be prevented. The latter can
be achieved by introducing different pinning centers such as
defects created by ion \cite{civale} or neutron irradiation
\cite{vandover}, lithographically introduced holes
\cite{hebard,VVM,VVM98}, or magnetic dots
\cite{otani,martin97,morgan,vanbael99}.\\ \indent Recently it was
proposed that vortices in a superconductor/ferromagnet (SC/FM)
multilayer are strongly pinned if the FM has a stripe domain
structure, which typically exists in thin magnetic films with
perpendicular anisotropy \cite{bula}. Enhanced pinning was
observed in SC/FM bilayers compared to SC reference films
\cite{garcia,zhang}. However, in these studies the domain
structure of the FM was unknown. In this Letter, we systematically
investigate the influence of the domains in a ferromagnetic Co/Pt
multilayer with perpendicular anisotropy on a {\em type-II}
superconducting Pb film grown on top of the FM. Stable domain
structures can be produced in a controlled way by magnetizing the
sample before measuring the superconducting properties of the Pb
film. We show that the strongest vortex pinning is obtained when
localized domains (bubble domains) are present in the FM, and that
by changing size and density of the bubbles one can control and
optimize the vortex pinning.\\ \indent The Co/Pt multilayer is
deposited on a Si substrate with amorphous SiO$_{2}$ top layer in
an MBE apparatus by e-beam evaporation. The multilayer has a
[Co(0.4~nm)/Pt(1.0~nm)]$_{10}$ structure on a 2.8~nm Pt base
layer. A 10~nm Ge film, a Pb film with thickness $d_{Pb}=50$~nm
and a 30~nm Ge capping layer are subsequently evaporated on the
Co/Pt multilayer (see Fig.~\ref{schema}) at a substrate
temperature of 77~K.
The Pb film has a critical temperature of $T_{c}=7.23$~K. The
penetration depth $\lambda(0) = 42$~nm and the coherence length
$\xi(0) = 41$~nm are estimated from measurements of the
temperature dependence of the upper critical field. The Ge film
between Pb and Co/Pt is insulating at low temperatures, so that
the proximity effects between Pb and Co/Pt are suppressed.\\
\indent The Co/Pt multilayer has perpendicular magnetic anisotropy
\cite{zeper}, which is confirmed by hysteresis loop measurements
using the magneto-optical Kerr effect (MOKE). Fig.~\ref{moke}
shows the magnetization $M_{fm}$ of the Co/Pt multilayer,
normalized to the saturation magnetization $M_{s}$, as a function
of the magnetic field $H$ applied perpendicular to the surface.
The loop has an almost rectangular shape with $H_{n} = 0.6$~kOe,
$H_{c} = 0.93$~kOe, and $H_{s} = 1.45$~kOe, where $H_{n}$, $H_{c}$
and $H_{s}$ are the nucleation, coercive and saturation field,
respectively.\\ \indent The microscopic domain structure of the
Co/Pt multilayer has been investigated by magnetic force
microscopy (MFM) at room temperature and $H=0$.
Figs.~\ref{mag}(a)-(c) show MFM images of the sample in different
stable remanent magnetic states.
In Fig.~\ref{mag}(a), the sample has been demagnetized by
oscillating $H$ around zero with decreasing amplitude
perpendicular to the surface. The dark and bright regions in the
image are domains with local magnetic moments $m$ either pointing
up ($m>0$) or down ($m<0$), perpendicular to the sample surface.
The observed {\em band domain structure} is typical for Co/Pt
multilayers \cite{valera}. The width of the band domains is about
$0.2-0.4$~$\mu$m.\\ \indent The images shown in Figs.~\ref{mag}(b)
and (c) are obtained after applying a negative field of $-10$~kOe,
sweeping $H$ to a positive value between $H_{n}$ and $H_{s}$ as
indicated in Fig.~\ref{moke}, and then removing $H$. When $H$ is
increased to a value slightly higher than $H_{n}$, small localized
domains (bubble domains) with magnetic moments $m>0$ nucleate in
the film. These bubbles are stable after removing $H$, as can be
seen in Fig.~\ref{mag}(b). When $H$ is further increased, the
bubbles grow and aggregate. At $H=H_{c}$, there is an equal number
of positive and negative $m$. If $H$ is further increased, the
number of negative $m$ decreases and the domains with $m<0$ become
localized (see Fig.~\ref{mag}(c)). At $H=H_{s}$, the film is
saturated with all $m>0$. We introduce the parameter $s$, which is
defined by the percentage of magnetic moments that are pointing up
($m>0$), to describe the different remanent magnetic states
obtained after this magnetization procedure. For the images shown
in Figs.~\ref{mag}(b) and (c), $s$ is given by the dark area
divided by the total area, which is in good agreement with $s$
calculated from Fig.~\ref{moke}.\\ \indent The bubble domains in
Fig.~\ref{mag}(b) typically have a size of about $0.35$~$\mu$m,
whereas the average size of the bubbles in Fig.~\ref{mag}(c) is
about $0.25$~$\mu$m. Another parameter changing during the
reversal process is the number of bubbles per unit area $\delta$.
For $H$ only slightly higher than $H_{n}$, $\delta$ is very small,
but once the bubbles have nucleated in the FM, the reversal
process is dominated by domain wall movement and $\delta$ does not
change significantly during this process. From the MFM images,
typical values of $\delta = 1-2$ bubbles/$\mu$m$^{2}$ are
derived.\\ \indent After MOKE and MFM measurements, the
superconducting Pb film is deposited on the Co/Pt multilayer and
the vortex pinning properties are studied by using a Quantum
Design SQUID magnetometer. First the FM is brought into a specific
magnetic state by applying the same magnetization procedure as for
the MFM measurements at 300~K. The magnetization of the FM
$M_{fm}$ is determined from $M(T)$ data as $M_{fm} = M (T>T_{c})$,
from which the parameter $s$ of each domain structure is
calculated. Repeating the $M(T)$ measurements in several applied
fields ($|H|<120$~Oe~$\ll H_{n}$, see Fig.~\ref{moke}) did not
change the value of $M_{fm}$, implying that the domain structures
are not altered by these small fields. The pinning strength
related to each domain structure is determined at $6.9$~K by
measuring $M$ as a function of $H$ applied perpendicular to the
surface from $-120$~Oe to +120~Oe. In order to compare the
magnetization curves measured in the different magnetic states,
$M_{fm}$ is subtracted from all curves shown in Fig.~\ref{mag}
(d)-(f), i.e.\ only the irreversible magnetization of the SC
$M_{sc} = M - M_{fm}$ is plotted. At $6.9~K$, $\xi(6.9 {\rm K}) =
0.19$~$\mu$m and the effective penetration depth $\Lambda(6.9 {\rm
K}) = \lambda^{2}(6.9 {\rm K})/d_{Pb} = 0.76$~$\mu$m. \\ \indent
The upper $M_{sc}(H)$ curve shown in Fig.~\ref{mag}(d) is obtained
after demagnetization (corresponding to the MFM image shown in
Fig.~\ref{mag}(a)) and the lower one in the $s=0$ state, i.e.\
with the FM fully magnetized in the negative direction. For both
magnetic states, a symmetric curve is found. The amplitude of
$M_{sc}(H)$ (which is a measure for the magnitude of $j_{c}(H)$
\cite{VVM98}) is lower for all $H$ when the FM is in the
demagnetized state compared to the $s=0$ state. This unambiguously
shows that the domain structure influences the SC, but does not
enhance the vortex pinning in this sample. In the $s=0$ state, the
stray field of the magnetic film has its largest amplitude at the
sample boundary and is much smaller above the center of the FM.
This means that the interior of the SC is only weakly influenced
by the stray field. Contrary to that, the alternating stray field
above the band domains suppresses the order parameter in the SC
and may even lead to the creation of vortex-antivortex pairs
connecting the domains with opposite magnetization and resulting
in a reduction of $j_{c}$ \cite{bula}. This is also reflected by a
decrease of $T_{c}$ of the SC to 7.04~K when the band domains are
present in the FM.\\ \indent When bubble domains are present in
the FM, the $M_{sc}(H)$ curve becomes strongly asymmetric with
respect to $H$. The direction of the asymmetry depends on the
mutual orientation of $H$ and the magnetic moments $m$ of the
bubbles. Fig.~\ref{mag}(e) shows two $M_{sc}(H)$ curves with $m$
of the bubbles pointing in the {\em positive direction} as in
Fig.~\ref{mag}(b). Both curves have a larger amplitude of $M_{sc}$
for positive H. The upper curve in Fig.~\ref{mag}(f) is obtained
after magnetizing the FM in $H= -20$~kOe~$\rightarrow +H_{c}
\rightarrow 0$, i.e.\ $s=0.5$. $M_{sc}(H)$ is almost completely
suppressed for this domain state. The other two curves of
Fig.~\ref{mag}(f) are obtained with $m$ of the bubbles in {\em
negative direction} as in Fig.~\ref{mag}(c). Also for this
configuration the $M_{sc}(H)$ curves are asymmetric, but reversed
with respect to the field polarity compared to
Fig.~\ref{mag}(e).\\ \indent The asymmetry of $M_{sc}(H)$ can be
explained by a magnetic interaction between vortices and bubble
domains. The magnetic interaction can be caused by (i) the
interaction between the magnetic field of the vortices $B$ and the
magnetic moments $m$ in the FM (interaction energy $E = - m B$,
resulting in attractive interaction for $m$ and $B$ having the
same polarity, and in repulsive interaction for $m$ and $B$ having
opposite polarity), and (ii) the interaction between the vortices
and the supercurrents generated in the SC by the stray field of
the domain structure. During the measurement of $M_{sc}(H)$ shown
in Fig.~\ref{mag}(e), the FM has had bubbles with $m>0$. Hence,
evaluating the interaction term $E = - m B$, the bubbles act as
pinning sites for vortices when $H>0$, resulting in the enhanced
amplitude of $M_{sc}$ for positive $H$. Consequently, the bubble
domains with $m<0$ in Fig.~\ref{mag}(f) are pinning centers for
vortices when $H<0$, giving a larger amplitude of $M_{sc}$ for
negative $H$. This behavior is consistent with the vortex pinning
properties of out-of-plane magnetized dots \cite{morgan}. Assuming
that one bubble can pin one vortex, we can estimate a "matching"
field $H_{match}$ at which all bubbles pin a vortex: $H_{match} =
\phi_{0} \delta \approx 20-40$~Oe, using $\delta = 1-2/\mu$m$^{2}$
and the flux quantum $\phi_{0}$. In this field range, $M_{sc}$
indeed has a strongly enhanced amplitude for $H>0$ in the $s=0.1$
state and for $H<0$ in the $s=0.85$ state, indicating strong
pinning of vortices.\\ \indent At small fields $|H|<3$~Oe, $j_{c}$
is not increased by any of the investigated domain structures
compared to the "reference" $s=0$ state in which domains are
absent. This can be attributed to the stray field above the
domains, which results in an overall reduction of the
superconducting order parameter $\psi$. Figs.~\ref{mag}(e) and (f)
also show that more effective vortex pinning is achieved for
bubbles with smaller size and lower density. The amplitude of
$M_{sc}$ can be increased up to a factor of 3 when comparing the
$s=0$ state with the lower curve of Fig.~\ref{mag}(f), and a
factor of 2 compared to the middle curve of Fig.~\ref{mag}(f).
This indicates that a reduced $\psi$ associated with bubbles with
larger size and higher density counteracts efficient vortex
pinning, even leading to the extremely weak amplitude of
$M_{sc}(H)$ for the $s=0.5$ state shown in the upper curve of
Fig.~\ref{mag}(f).\\ \indent In conclusion, we have shown that the
domain structure in a magnetic film can significantly influence
the vortex pinning in a type-II superconductor. By tuning the
density of bubble domains in the magnetic film, one can control
the vortex pinning. This could be used to design logical devices
based on the control of superconductivity by magnetic domain
structures \cite{bula}.\\ \indent The authors acknowledge L. Van
Look, K. Temst and G. G\"untherodt for help with sample
preparation, and J. Swerts for MOKE measurements. This work was
supported by the Belgian IUAP and the Flemish GOA programs, by the
ESF "VORTEX" program and by the Fund for Scientific Research
(F.W.O.) - Flanders. MJVB is a Postdoctoral Research Fellow of the
F.W.O. - Flanders.


\begin{thebibliography}{14}
\expandafter\ifx\csname
natexlab\endcsname\relax\def\natexlab#1{#1}\fi
\expandafter\ifx\csname bibnamefont\endcsname\relax
  \def\bibnamefont#1{#1}\fi
\expandafter\ifx\csname bibfnamefont\endcsname\relax
  \def\bibfnamefont#1{#1}\fi
\expandafter\ifx\csname citenamefont\endcsname\relax
  \def\citenamefont#1{#1}\fi
\expandafter\ifx\csname url\endcsname\relax
  \def\url#1{\texttt{#1}}\fi
\expandafter\ifx\csname
urlprefix\endcsname\relax\def\urlprefix{URL }\fi
\providecommand{\bibinfo}[2]{#2}
\providecommand{\eprint}[2][]{\url{#2}}

\bibitem[{\citenamefont{Civale et~al.}(1991)\citenamefont{Civale, Marwick, K.,
  Kirk, Thompson, Krusinelbaum, Sun, Clem, and Holtzberg}}]{civale}
\bibinfo{author}{\bibfnamefont{L.}~\bibnamefont{Civale}},
  \bibinfo{author}{\bibfnamefont{A.~D.} \bibnamefont{Marwick}},
  \bibinfo{author}{\bibfnamefont{T.}~\bibnamefont{K.}},
  \bibinfo{author}{\bibfnamefont{M.~A.} \bibnamefont{Kirk}},
  \bibinfo{author}{\bibfnamefont{J.~R.} \bibnamefont{Thompson}},
  \bibinfo{author}{\bibfnamefont{L.}~\bibnamefont{Krusinelbaum}},
  \bibinfo{author}{\bibfnamefont{Y.}~\bibnamefont{Sun}},
  \bibinfo{author}{\bibfnamefont{J.~R.} \bibnamefont{Clem}}, \bibnamefont{and}
  \bibinfo{author}{\bibfnamefont{F.}~\bibnamefont{Holtzberg}},
  \bibinfo{journal}{Phys. Rev. Lett.} \textbf{\bibinfo{volume}{67}},
  \bibinfo{pages}{648} (\bibinfo{year}{1991}).

\bibitem[{\citenamefont{vanDover et~al.}(1989)\citenamefont{vanDover, Gyorgy,
  Schneemeyer, Mitchell, Rao, Puzniak, and Waszczak}}]{vandover}
\bibinfo{author}{\bibfnamefont{R.~B.} \bibnamefont{vanDover}},
  \bibinfo{author}{\bibfnamefont{E.~M.} \bibnamefont{Gyorgy}},
  \bibinfo{author}{\bibfnamefont{L.~F.} \bibnamefont{Schneemeyer}},
  \bibinfo{author}{\bibfnamefont{J.~W.} \bibnamefont{Mitchell}},
  \bibinfo{author}{\bibfnamefont{K.~V.} \bibnamefont{Rao}},
  \bibinfo{author}{\bibfnamefont{R.}~\bibnamefont{Puzniak}}, \bibnamefont{and}
  \bibinfo{author}{\bibfnamefont{J.~V.} \bibnamefont{Waszczak}},
  \bibinfo{journal}{Nature} \textbf{\bibinfo{volume}{342}}, \bibinfo{pages}{55}
  (\bibinfo{year}{1989}).

\bibitem[{\citenamefont{Fiory et~al.}(1978)\citenamefont{Fiory, Hebard, and
  Somekh}}]{hebard}
\bibinfo{author}{\bibfnamefont{A.~T.} \bibnamefont{Fiory}},
  \bibinfo{author}{\bibfnamefont{A.~F.} \bibnamefont{Hebard}},
  \bibnamefont{and} \bibinfo{author}{\bibfnamefont{S.}~\bibnamefont{Somekh}},
  \bibinfo{journal}{Appl. Phys. Lett.} \textbf{\bibinfo{volume}{32}},
  \bibinfo{pages}{73} (\bibinfo{year}{1978}).

\bibitem[{\citenamefont{Moshchalkov et~al.}(1999)\citenamefont{Moshchalkov,
  Bruyndoncx, Van~Look, Van~Bael, and Tonomura}}]{VVM}
\bibinfo{author}{\bibfnamefont{V.~V.} \bibnamefont{Moshchalkov}},
  \bibinfo{author}{\bibfnamefont{V.}~\bibnamefont{Bruyndoncx}},
  \bibinfo{author}{\bibfnamefont{L.}~\bibnamefont{Van~Look}},
  \bibinfo{author}{\bibfnamefont{M.~J.} \bibnamefont{Van~Bael}},
  \bibnamefont{and} \bibinfo{author}{\bibfnamefont{A.}~\bibnamefont{Tonomura}},
  in \emph{\bibinfo{booktitle}{Handbook of Nanostructured Materials and
  Nanotechnology}}, edited by \bibinfo{editor}{\bibfnamefont{H.~S.}
  \bibnamefont{Nalwa}} (\bibinfo{publisher}{Academic Press},
  \bibinfo{address}{San Diego}, \bibinfo{year}{1999}),
  vol.~\bibinfo{volume}{3}, chap.~\bibinfo{chapter}{9}, p.
  \bibinfo{pages}{451}.

\bibitem[{\citenamefont{Moshchalkov et~al.}(1998)\citenamefont{Moshchalkov,
  Baert, Metlushko, Rosseel, Van~Bael, Temst, Bruynseraede, and
  Jonckheere}}]{VVM98}
\bibinfo{author}{\bibfnamefont{V.~V.} \bibnamefont{Moshchalkov}},
  \bibinfo{author}{\bibfnamefont{M.}~\bibnamefont{Baert}},
  \bibinfo{author}{\bibfnamefont{V.~V.} \bibnamefont{Metlushko}},
  \bibinfo{author}{\bibfnamefont{E.}~\bibnamefont{Rosseel}},
  \bibinfo{author}{\bibfnamefont{M.~J.} \bibnamefont{Van~Bael}},
  \bibinfo{author}{\bibfnamefont{K.}~\bibnamefont{Temst}},
  \bibinfo{author}{\bibfnamefont{Y.}~\bibnamefont{Bruynseraede}},
  \bibnamefont{and}
  \bibinfo{author}{\bibfnamefont{R.}~\bibnamefont{Jonckheere}},
  \bibinfo{journal}{Phys. Rev. B} \textbf{\bibinfo{volume}{57}},
  \bibinfo{pages}{3615} (\bibinfo{year}{1998}).

\bibitem[{\citenamefont{Otani et~al.}(1993)\citenamefont{Otani, Pannetier,
  Nozi{\`e}res, and Givord}}]{otani}
\bibinfo{author}{\bibfnamefont{Y.}~\bibnamefont{Otani}},
  \bibinfo{author}{\bibfnamefont{B.}~\bibnamefont{Pannetier}},
  \bibinfo{author}{\bibfnamefont{J.~P.} \bibnamefont{Nozi{\`e}res}},
  \bibnamefont{and} \bibinfo{author}{\bibfnamefont{D.}~\bibnamefont{Givord}},
  \bibinfo{journal}{J. Magn. Magn. Mat.} \textbf{\bibinfo{volume}{126}},
  \bibinfo{pages}{622} (\bibinfo{year}{1993}).

\bibitem[{\citenamefont{Mart{\'\i}n et~al.}(1997)\citenamefont{Mart{\'\i}n,
  V\'elez, Nogu\'es, and Schuller}}]{martin97}
\bibinfo{author}{\bibfnamefont{J.~I.} \bibnamefont{Mart{\'\i}n}},
  \bibinfo{author}{\bibfnamefont{M.}~\bibnamefont{V\'elez}},
  \bibinfo{author}{\bibfnamefont{J.}~\bibnamefont{Nogu\'es}}, \bibnamefont{and}
  \bibinfo{author}{\bibfnamefont{I.~K.} \bibnamefont{Schuller}},
  \bibinfo{journal}{Phys. Rev. Lett.} \textbf{\bibinfo{volume}{79}},
  \bibinfo{pages}{1929} (\bibinfo{year}{1997}).

\bibitem[{\citenamefont{Morgan and Ketterson}(1998)}]{morgan}
\bibinfo{author}{\bibfnamefont{D.~J.} \bibnamefont{Morgan}} \bibnamefont{and}
  \bibinfo{author}{\bibfnamefont{J.~B.} \bibnamefont{Ketterson}},
  \bibinfo{journal}{Phys. Rev. Lett.} \textbf{\bibinfo{volume}{80}},
  \bibinfo{pages}{3614} (\bibinfo{year}{1998}).

\bibitem[{\citenamefont{Van~Bael et~al.}(1999)\citenamefont{Van~Bael, Temst,
  Moshchalkov, and Bruynseraede}}]{vanbael99}
\bibinfo{author}{\bibfnamefont{M.~J.} \bibnamefont{Van~Bael}},
  \bibinfo{author}{\bibfnamefont{K.}~\bibnamefont{Temst}},
  \bibinfo{author}{\bibfnamefont{V.~V.} \bibnamefont{Moshchalkov}},
  \bibnamefont{and}
  \bibinfo{author}{\bibfnamefont{Y.}~\bibnamefont{Bruynseraede}},
  \bibinfo{journal}{Phys. Rev. B} \textbf{\bibinfo{volume}{59}},
  \bibinfo{pages}{14674} (\bibinfo{year}{1999}).

\bibitem[{\citenamefont{Bulaevskii et~al.}(2000)\citenamefont{Bulaevskii,
  Chudnovsky, and Maley}}]{bula}
\bibinfo{author}{\bibfnamefont{L.~N.} \bibnamefont{Bulaevskii}},
  \bibinfo{author}{\bibfnamefont{E.~M.} \bibnamefont{Chudnovsky}},
  \bibnamefont{and} \bibinfo{author}{\bibfnamefont{M.~P.} \bibnamefont{Maley}},
  \bibinfo{journal}{Appl. Phys. Lett.} \textbf{\bibinfo{volume}{76}},
  \bibinfo{pages}{2594} (\bibinfo{year}{2000}).

\bibitem[{\citenamefont{Garc{\'\i}a-Santiago
  et~al.}(2000)\citenamefont{Garc{\'\i}a-Santiago, S\'{a}nchez, Varela, and
  Tejada}}]{garcia}
\bibinfo{author}{\bibfnamefont{A.}~\bibnamefont{Garc{\'\i}a-Santiago}},
  \bibinfo{author}{\bibfnamefont{F.}~\bibnamefont{S\'{a}nchez}},
  \bibinfo{author}{\bibfnamefont{M.}~\bibnamefont{Varela}}, \bibnamefont{and}
  \bibinfo{author}{\bibfnamefont{J.}~\bibnamefont{Tejada}},
  \bibinfo{journal}{Appl. Phys. Lett.} \textbf{\bibinfo{volume}{77}},
  \bibinfo{pages}{2900} (\bibinfo{year}{2000}).

\bibitem[{\citenamefont{Zhang et~al.}(2001)\citenamefont{Zhang, Wen, Zheng,
  Xiong, and Lian}}]{zhang}
\bibinfo{author}{\bibfnamefont{X.~X.} \bibnamefont{Zhang}},
  \bibinfo{author}{\bibfnamefont{G.~H.} \bibnamefont{Wen}},
  \bibinfo{author}{\bibfnamefont{R.~K.} \bibnamefont{Zheng}},
  \bibinfo{author}{\bibfnamefont{G.~C.} \bibnamefont{Xiong}}, \bibnamefont{and}
  \bibinfo{author}{\bibfnamefont{G.~J.} \bibnamefont{Lian}},
  \bibinfo{journal}{Europhys. Lett.} \textbf{\bibinfo{volume}{56}},
  \bibinfo{pages}{119} (\bibinfo{year}{2001}).

\bibitem[{\citenamefont{Zeper et~al.}(1989)\citenamefont{Zeper, Greidanus,
  Carcia, and Fincher}}]{zeper}
\bibinfo{author}{\bibfnamefont{W.~B.} \bibnamefont{Zeper}},
  \bibinfo{author}{\bibfnamefont{F.~J.~A.} \bibnamefont{Greidanus}},
  \bibinfo{author}{\bibfnamefont{P.~F.} \bibnamefont{Carcia}},
  \bibnamefont{and} \bibinfo{author}{\bibfnamefont{C.~R.}
  \bibnamefont{Fincher}}, \bibinfo{journal}{J. Appl. Phys.}
  \textbf{\bibinfo{volume}{65}}, \bibinfo{pages}{4971} (\bibinfo{year}{1989}).

\bibitem[{\citenamefont{Valera et~al.}(1995)\citenamefont{Valera, Farley, Hoon,
  Zhou, Mcvitie, and Chapman}}]{valera}
\bibinfo{author}{\bibfnamefont{M.~S.} \bibnamefont{Valera}},
  \bibinfo{author}{\bibfnamefont{A.~N.} \bibnamefont{Farley}},
  \bibinfo{author}{\bibfnamefont{S.~R.} \bibnamefont{Hoon}},
  \bibinfo{author}{\bibfnamefont{L.}~\bibnamefont{Zhou}},
  \bibinfo{author}{\bibfnamefont{S.}~\bibnamefont{Mcvitie}}, \bibnamefont{and}
  \bibinfo{author}{\bibfnamefont{J.~N.} \bibnamefont{Chapman}},
  \bibinfo{journal}{Appl. Phys. Lett.} \textbf{\bibinfo{volume}{67}},
  \bibinfo{pages}{2566} (\bibinfo{year}{1995}).

\end{thebibliography}
\end{document}